\documentclass[trackchanges, twocolumn]{aastex701}

\usepackage{acro}
\usepackage{amsmath}
\usepackage{xspace}
\usepackage{float}

\acsetup{
  single    = false,
  list/sort  = true,
  cite/group = true,
  cite/group/cmd = \citealt,
  cite/group/pre = {\xspace; }
}

\DeclareAcronym{adc}{
  short = ADC,
  long = analog-to-digital converter,
}
\DeclareAcronym{edm}{
  short = EDM,
  long = effective detector model,
}

\DeclareAcronym{inl}{
  short = INL,
  long = integral nonlinearity,
}

\DeclareAcronym{jwst}{
  short = JWST,
  long = \textit{James Webb Space Telescope},
  cite = {Gardner2023}
}

\DeclareAcronym{ers}{
  short = ERS,
  long = Early Release Science
}

\DeclareAcronym{niriss}{
  short = NIRISS,
  long = Near Infrared Imager and Slitless Spectrograph,
  cite={niriss2023}
}

\DeclareAcronym{ls}{
  short = LS periodogram,
  long = Lomb-Scargle Periodogram,
  cite={Lomb1976,Scargle1982,VanderPlas2018}
}

\DeclareAcronym{psf}{
  short = PSF,
  long = point spread function
}

\DeclareAcronym{mast}{
  short = MAST,
  long = Barbara A. Mikulski Archive for Space Telescopes
}

\DeclareAcronym{soss}{
  short = SOSS,
  long = Single Object Slitless Spectroscopy
}

\DeclareAcronym{bfe}{
  short = BFE,
  long = brighter-fatter effect,
  cite = {Antilogus2014,Hirata2020}
}

\DeclareAcronym{ami}{
  short = AMI,
  long = Aperture Masking Interferometer,
  cite = {Sivaramakrishnan2023}
}

\DeclareAcronym{wfss}{
  short = WFSS,
  long = Wide Field Slitless Spectroscopy
}

\DeclareAcronym{h2rg}{
  short = H2RG,
  long = Hawaii-2RG,
  cite = {Blank2011}
}

\DeclareAcronym{adu}{
  short = ADU,
  long = analog-to-digital units,
}

\DeclareAcronym{nuts}{
  short = NUTS,
  long = No-U-Turn Sampler,
}
\submitjournal{PASP}

\begin{document}

\title[ADC Nonlinearity in JWST/NIRISS]{Calibration of an Analog-to-Digital Conversion Nonlinearity in JWST/NIRISS}

\author[orcid=0000-0001-9145-8444]{Shashank Dholakia}
\affiliation{School of Mathematics and Physics, University of Queensland, St Lucia, QLD 4072, Australia}
\email[show]{s.dholakia@uq.edu.au}

\author[orcid=0000-0001-6263-4437]{Shishir Dholakia}
\affiliation{Centre for Astrophysics, University of Southern Queensland, West Street, Toowoomba, QLD 4350, Australia}
\email{dholakia.shishir@gmail.com}

\author[orcid=0000-0003-2595-9114]{Benjamin~J.~S.~Pope}
\affiliation{School of Mathematical and Physical Sciences, Macquarie University, Sydney, NSW 2109, Australia}
\affiliation{Astrophysics and Space Technologies Research Centre, Macquarie University, Sydney, NSW 2109, Australia}
\email{benjamin.pope@mq.edu.au}

\author[orcid=0000-0002-1015-9029]{Louis~Desdoigts}
\affiliation{Leiden Observatory, Niels Bohrweg 2, Leiden 2300RA, The Netherlands}
\affiliation{Sydney Institute for Astronomy, School of Physics, University of Sydney, NSW 2006, Australia}
\email{desdoigts@strw.leidenuniv.nl}

\author[orcid=0000-0003-2259-3911]{Shrishmoy Ray}
\affiliation{School of Mathematical \& Physical Sciences, Macquarie University, 12 Wally's Walk, Macquarie Park, NSW 2113}
\email{shrishmoy.ray@mq.edu.au}

\author[orcid=0000-0001-7026-6291]{Peter~G.~Tuthill}
\affiliation{Sydney Institute for Astronomy, School of Physics, University of Sydney, NSW 2006, Australia}
\email{peter.tuthill@sydney.edu.au}

\author[orcid=0000-0003-1251-4124]{Anand~Sivaramakrishnan}
\affiliation{Space Telescope Science Institute, 3700 San Martin Drive, Baltimore, MD 21218, USA}
\affiliation{Astrophysics Department, American Museum of Natural History, 79th ST at CPW, New York, NY, 10024, USA}
\affiliation{Department of Physics and Astronomy, Johns Hopkins University, 3701 San Martin Drive, Baltimore, MD 21218, USA}
\email{anand@stsci.edu}

\begin{abstract}
We quantify an unusual flux-dependent systematic which is periodic in raw counts in flight data from the \textit{James Webb Space Telescope's} Near Infrared Imager and Slitless Spectrograph (JWST/NIRISS), used extensively for exoplanet imaging and spectroscopy. Originally discovered in the aperture masking interferometry (AMI) mode, it also manifests in the Single Object Slitless Spectroscopy (SOSS) mode with the same dominant period of 1024 in raw analog-to-digital units (ADU). The likely cause of the signal is an analog-to-digital converter (ADC) integral nonlinearity (INL) in which case it will apply to all observations taken with the NIRISS instrument. Fortunately, it is straightforward to correct the data in postprocessing. The periodic INL is shown to be flux-dependent, increasing in amplitude with higher pixel counts on the detector. We derive a model of this periodic INL by fitting a combination of a polynomial and sinusoid multiplied with the residuals of ramp fits to the uncalibrated data and find an amplitude of 125\,ppm, up to a 2.5-count shift for a pixel with 20,000\,ADU. We apply this model to correct the well-studied NIRISS SOSS Program ERS~1366 dataset of WASP-39\,b and reduce the data into a transmission spectrum. We find that our corrected transmission spectrum removes the INL systematic from the uncorrected spectrum at the 30\,ppm level across both orders, and also corrects a 55\,ppm offset between Order 1 and Order 2. We recommend a larger scale data-driven calibration of the periodic INL and the adoption of the outcome into NIRISS data pipelines. 

\end{abstract}


\section{Introduction}
\label{sec:intro}


The \acf{jwst}  is a revolutionary flagship 6\,m diameter infrared observatory built by NASA/ESA/CSA and launched in 2021. It has exceeded expectations across observing modes \citep{Rigby2023} and is becoming a workhorse of scientific breakthroughs from exoplanets to cosmology.

The \acf{niriss} is one of the four science instruments on board the \acl{jwst}, and has modes for imaging and spectroscopy, both with the \acf{wfss} and \acf{soss} modes \citep{albert2023}. \ac{niriss} has been used widely for transmission spectroscopy \citep[e.g.,][]{feinstein2023, holmberg2023, radica2023, cadieux2024, fournier2024, ahrer2025, fournier2025, Taylor2025} and offers the only dedicated interferometric mode in \ac{jwst}, the \acf{ami}.

As it is a \textit{slitless} spectrograph, the reduction and processing of the raw data presents a challenge, including characterization of background sources, ghosts, and other such effects which could contaminate the data. Much attention has been devoted to the characterization of such effects and the reduction of other systematics in \ac{wfss} \citep{willott2022, hviding2024}, \ac{ami} \citep{sallum2024, amigo2025} and \ac{soss} \citep{baines2023, radica2023, holmberg2023, radica2024, baines2024, fournier2024}. 

In the first commissioning and \ac{ers} publications using \ac{ami}, the first analyses did not initially reach photon-limited, diffraction-limited performance \citep{sallum2024,Ray2025}. A consensus quickly developed that this was a consequence of charge migration, otherwise known as the \acf{bfe}. This is a detector nonlinearity in which signal from bright pixels blurs into neighbours, so that the overall effect is for the \acp{psf} to appear ``fatter''. This turned out to be ruinous for the assumptions of linearity that normally underpin the calibration of interferometric data, introducing systematics into observables recovered from interferograms sampled by the instrument's \acf{h2rg} detector.

\subsection{The ADC Nonlinearity}
In the course of building an end-to-end differentiable model of the instrument to predict and invert the \ac{bfe}-affected PSF \citep{Desdoigts2024, amigo2025}, an unusual systematic became apparent: a residual that was periodic \textit{in raw data units}, in all pixels, with a period of 1024 counts. 

Such systematics manifest not in time or space but in raw counts, at a period of a power of two, immediately suggests the \ac{adc} which converts voltages on the pixel outputs into a digital signal that can be stored, downlinked, and processed. It is common for \acp{adc} to have an \ac{inl}, which is an offset between the reported digital output and the ideal linear output. It is an ``integral'' signal because it represents the accumulated difference from ideal performance of a differential nonlinearity, which is the early or late flipping of each binary digitized signal level. These \acp{inl} frequently therefore contain periods at powers of 2, corresponding to the stages of amplification and binary digitization in the \ac{adc} hardware. 

In references about the \ac{inl} of the \ac{h2rg} detectors on \ac{jwst} and their SIDECAR readout electronics \citep{Loose2005,Chen2014}, we see examples of qualitatively similar curves in amplitude and period for other SIDECAR instances. An \ac{inl} at 1024 period is a possible consequence of low amplifier power in a successive approximation ADC; a full explanation of the hardware origin of this \ac{inl} is beyond the scope of this paper as the relevant information is subject to commercial and security based restrictions and unavailable to the authors.

In this paper we go beyond the initial exploration presented in \citet{amigo2025}, and characterize this \ac{inl} in other \ac{niriss} data. We propose an algorithm for iteratively fitting this \ac{inl} to a large volume of ramp-level \ac{soss} data in Section~\ref{sec:inl}, and then apply this to a well-studied transmission spectroscopy dataset in Section~\ref{sec:wasp39}. We provide open source code implementing our correction, \href{https://github.com/shashankdholakia/niriss-cal-inl}{available on GitHub}.

Although not presented in this paper, it may be reassuring to users of \ac{jwst}'s other instruments that we have searched and been unable to find any comparably strong nonlinear systematic in data from NIRCam or NIRSpec, both of which use similar \ac{h2rg} detectors and \ac{adc} electronics.

\section{Inferring the Integral nonlinearity} \label{sec:inl}

We use the raw, Stage~0 \ac{niriss}/\ac{soss} data on WASP-121\,b from GTO~1201 (PI Lafrenière) in order to infer the existence of the \ac{inl} and learn its characteristics. The instrument configuration for this observation uses the SUBSTRIP256 array with $256\times2048$ pixels and 6 groups per integration \citep{splinter2025}. We use the first 188 out of the total 3452 integrations (the exposure is broken into files with 188 integrations each) to reduce the computational expense and flux variations of the transiting planet over the observations.

We start by constructing a mask to prevent low or high signal regions from inclusion in the analysis. Pixels with pedestal counts above 30,000 ADU are first excluded to avoid saturation. We then exclude pixels which have accumulated less than 300 counts over the 6 groups, which eliminates regions outside the spectral trace with very low signal. Finally, we restrict further to include only pixels with a zero-group count greater than $10,000$ and a final group count less than $20,000$. This was chosen as a narrow range in count space where the group level data can be well approximated as linear. 

\begin{figure*}
    \centering
    \includegraphics[width=\linewidth]{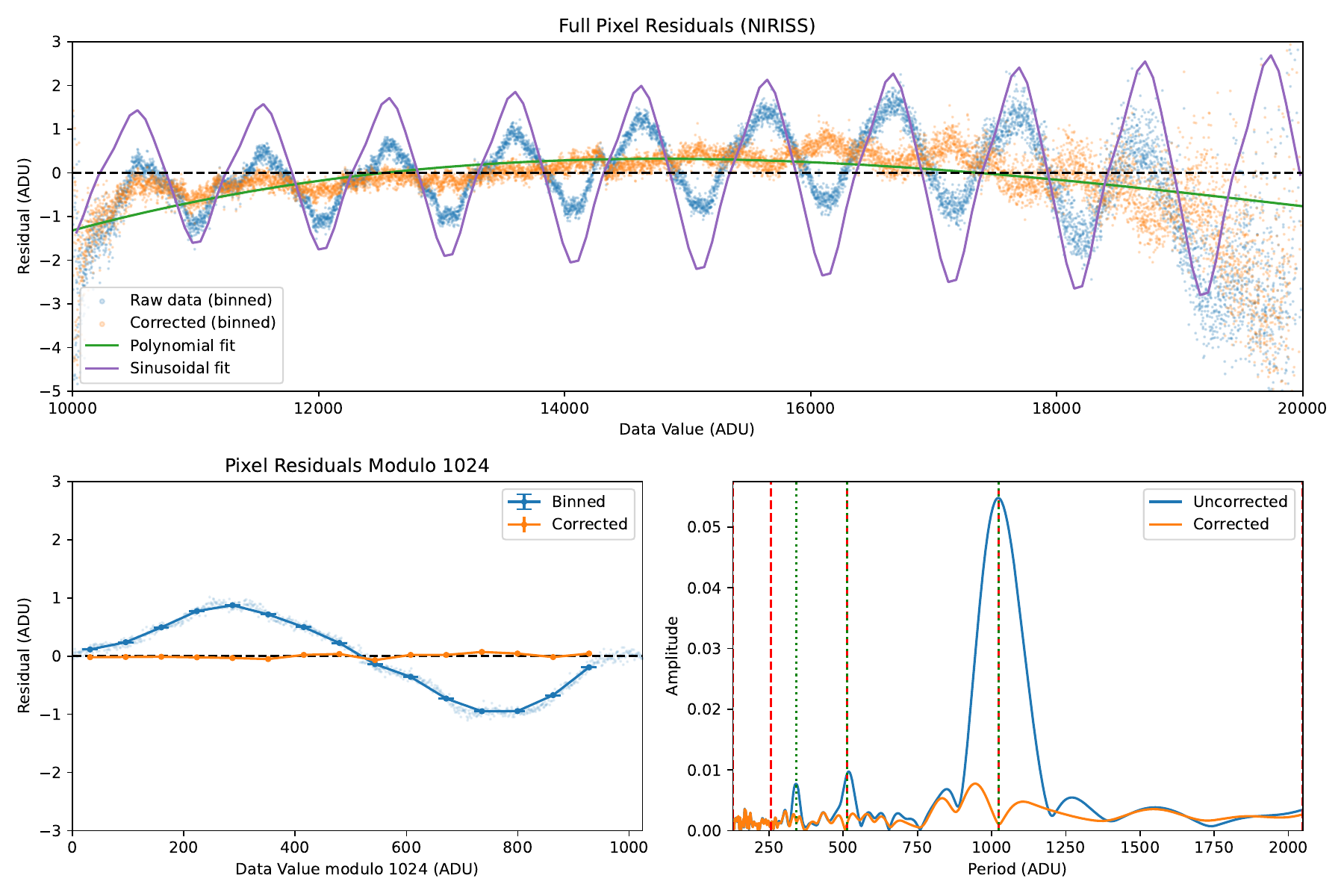}
    \caption{Results from the inference of the integral nonlinearity using Stage~0 NIRISS/SOSS data on WASP-121\,b. Residuals to the ramp fitting stage (blue points, top) show a periodic anomaly with a period of roughly 1024 counts. Fitting a nonlinearity model with a polynomial plus a sinusoid proportional to the signal level (green, purple, respectively) to the residuals reduces the sinusoidal anomaly. The corrected data is shown in orange, where only the sinusoidal component has been removed. Lower left panel shows the same data folded by the dominant 1024 period and binned (uncorrected in blue and corrected in orange). Bottom right panel shows the Lomb-Scargle periodogram of the ramp fit residuals against the raw count for each pixel (blue) showing the same dominant peak at 1024 ADU. We show in red each power of 2, and in green the harmonics of the dominant 1024 period.}
    \label{fig:adcinl}
\end{figure*}

We then perform a linear ramp fit over the 6 groups to each of the 188 integrations on all the remaining pixels in the array. Comparing the residuals of this ramp fit against the respective pixel values, we find a high frequency artifact, shown in Figure~\ref{fig:adcinl}. We flatten the data cube such that each pixel in each integration and group over the detector area has a readout data value (in ADU, plotted on the x-axis in the top panel) and a residual from the ramp fit (also in ADU, plotted on the y-axis in the top panel). Binning these data (shown with the blue scatter points) shows a clear sinusoidal pattern with the dominant 1024 period. Furthermore, this pattern evolves up in amplitude with increasing readout value. As such, we adopt a multiplicative model for this \ac{inl} where the amplitude is a fraction of the raw pixel value.

Taking the \ac{ls} of each pixel's ramp fit residual versus its raw count value, there is a dominant peak at a period of 1024, and smaller peaks at other powers of two and harmonics of 1024. The dominant period of $2^{10}$ is consistent with an issue originating with the \ac{adc}. We show the \ac{ls} of the ramp fit residuals against the raw pixel count in the lower right panel of Figure~\ref{fig:adcinl}.


We use an iterative approach to infer a calibration for the periodic \acl{inl} that accounts for other detector nonlinearity. We define a model for the pixel charge on the detector as:

\begin{equation}
q = f \cdot (g \cdot t_g)
\end{equation}
where q is the pixel charge (in electrons $e^{-}$), $f$ is the flux in $e^{-}/$s, $g$ is the integer group number and $t_g$ is the integration time in seconds for each group. The charge is then read; we model the idealized, linear, count level (in data numbers $\mathrm{DN}$, or \ac{adu}) (without any nonlinearities) after this read as:
\begin{equation}
C_{\mathrm{lin}} = \gamma + \beta \cdot q = \gamma + \beta f g
\end{equation}
where $\gamma$ is the pedestal value (ADU) and $\beta^{-1}$ is the gain ($e^{-}/\mathrm{DN}$). In practice, the count level after the read is affected by nonlinearities, including the periodic \ac{inl}. We model these nonlinearities as a global, multiplicative effect shared across all pixels consisting of a polynomial term to model classical nonlinearity and a sinusoid representing the periodic \ac{inl}. The readout value after the \ac{adc} is therefore modelled as:
\begin{equation} \label{eq:nonlinearity}
r = C_{\mathrm{lin}} (1 + N_{\text{poly}}(x) + N_{\text{INL}}(C_{\text{lin}})).
\end{equation} 

\noindent We use a quartic polynomial without a constant term (which would be degenerate with the pedestal value) as follows:
\begin{equation} \label{eq:polynomial}
N_{\text{poly}}(x) = a x + b x^2 + c x^3 + d x^4
\end{equation}
where $x$ is a normalized count value (in DN) $x = \frac{C_{\mathrm{lin}} - C_{\text{mid}}}{C_{\text{scale}}}$. We choose a quartic polynomial as a balance to fit for classical nonlinearities in the ramp while remaining smooth enough to avoid fitting out the periodic \ac{inl}.

Finally, the periodic \ac{inl} itself is modelled as a sinusoid:

\begin{equation} \label{eq:sinusoid}
N_{\text{INL}}(C_{\text{lin}}) = \sum_{k} \left[ a_k \sin\left(\frac{2\pi C_{\text{lin}}}{P_k}\right) + b_k \cos\left(\frac{2\pi C_{\text{lin}}}{P_k}\right) \right].
\end{equation}

The periods for the sinusoid representing the \ac{inl} are assumed from first principles to occur at powers of 2 or integer harmonics thereof. The highest peaks in the periodogram in Figure~\ref{fig:adcinl} correspond to periods of 1024, 512 and 1024/3 ADU, corroborating this. We attempt to iteratively learn the corresponding Fourier coefficients of these 3 peaks as shown in Eq.~\ref{eq:sinusoid} as described below by fitting Eq.~\ref{eq:nonlinearity}. We fit a linear ramp over the 6 groups to infer the count rate $\beta f$ ($\mathrm{DN}/s$) and pedestal values $\gamma$ for each of the 188 integrations. Starting with an initial amplitude of 0 for each, we first subtract the polynomial term in Eq~\ref{eq:polynomial} and a 6-term sinusoid in Eq.~\ref{eq:sinusoid} before performing another linear ramp fit. After this step, the polynomial coefficients and sinusoidal amplitude are updated from the residuals. We find that without this iterative procedure, each individual ramp fit is strongly aliased with respect to the sinusoid and hence the ramp fit results in biased slopes and amplitudes. 

The pedestal $\gamma$ and count rate $\beta f$ are fit in the linear ramps for each integration individually for each pixel on the detector. The fitting procedure treats the polynomial coefficients in Eq.~\ref{eq:polynomial} and the periodic \ac{inl} coefficients in Eq.~\ref{eq:sinusoid} as free parameters but constant across the detector. The readout value $r$ and the normalized count value $x$ are taken from the raw data and we calculate the ideal linear count $C_{lin}$ using the other parameters. The pixel gains and sinusoid period are fixed. Our fitting code is written in \textsc{Jax} using a block coordinate descent approach to mitigate computational and memory limitations. The total RMS value of the ramp fits starts at $20.6$ at the initial fit, where coefficients for both the polynomial and sinusoid are 0, and we continue iterations until the RMS changes by no more than $10^{-7}$, resulting in an RMS of $18.4$ after 28 iterations.

We visualize the periodic \ac{inl} along with the results of the fitting procedure in Figure~\ref{fig:adcinl} and show the best-fit coefficients in Table~\ref{tab:coeffs}. The ramp fit residuals across all pixels and integrations display a clear sinusoidal pattern that evolves in amplitude with the data value. The resulting fit for the polynomial plus sinusoidal nonlinearity removes much of the periodicity, although some residuals are left, especially towards the lower and higher range of the signal, possibly due to a more complex model for the amplitude than our assumption of a linear increase over data numbers. We discuss potential sources for the remaining periodic signal towards the end of Section~\ref{sec:discussion}. In particular, throughout the rest of this work we assume the sinusoidal \ac{inl} inferred through this procedure remains constant for different datasets with varying groups, integrations, and readout patterns. This assumption is validated by comparison to \citet{amigo2025}, which shows a similar amplitude effect in the \ac{ami} mode as well as further in Section~\ref{sec:wasp39}, where we assess a correction based on the sinusoid derived above in this Section using a different, well-characterized \ac{niriss} dataset.

\begin{table}[htbp]
\centering
\begin{minipage}{\columnwidth}
    \centering
    \begin{tabular}{lcc}
    \hline \hline
    Period & $a_k$ & $b_k$ \\
    \hline \hline
    $1024/3$ & -0.64  & 4.13  \\
    $512$    & -2.78  & 4.83  \\
    $1024$   & 124.34 & -6.41 \\
    \hline
    \end{tabular}
    \vspace{0.2cm} 
    \caption{Best-fit sinusoidal coefficients $a_k$ and $b_k$ from Eq.~\ref{eq:sinusoid} (in ppm) for each period included in the fit.}
    \label{tab:coeffs}
\end{minipage}
\end{table}

\section{Application of the Calibration to WASP-39\,b Transmission Spectrum} \label{sec:wasp39}


To test the \ac{adc} \acl{inl} learned in Section~\ref{sec:inl}, we benchmark on the \ac{jwst} \ac{ers} program data on WASP-39\,b. As part of the Early Release Science program for JWST, this dataset is among the best characterized, and provides a good testbed for evaluating the effects of correcting for the periodic \acl{inl}. We start with Stage~0 data as in Section~\ref{sec:inl}, except we include the entire dataset on WASP-39\,b totaling 537 integrations. We then create an identical copy of the Stage~0 data, except where we have divided out only the sinusoidal component of the nonlinearity as inferred above (Section~\ref{sec:inl}). We use the same sinusoid coefficients as for the previous dataset and simple extrapolation to handle values outside the range probed by the fit. Other sources of nonlinearity (which we model as a quartic polynomial) are already accounted for in the pipeline procedures described below; we assume the periodic \ac{inl} is the only unknown and uncharacterized effect. We designate this copy as the INL `corrected' dataset, in contrast to the original Stage~0 data (`uncorrected') where no modifications have been made. 

\begin{figure*}[h!]
    \centering
    \includegraphics[width=0.9\linewidth]{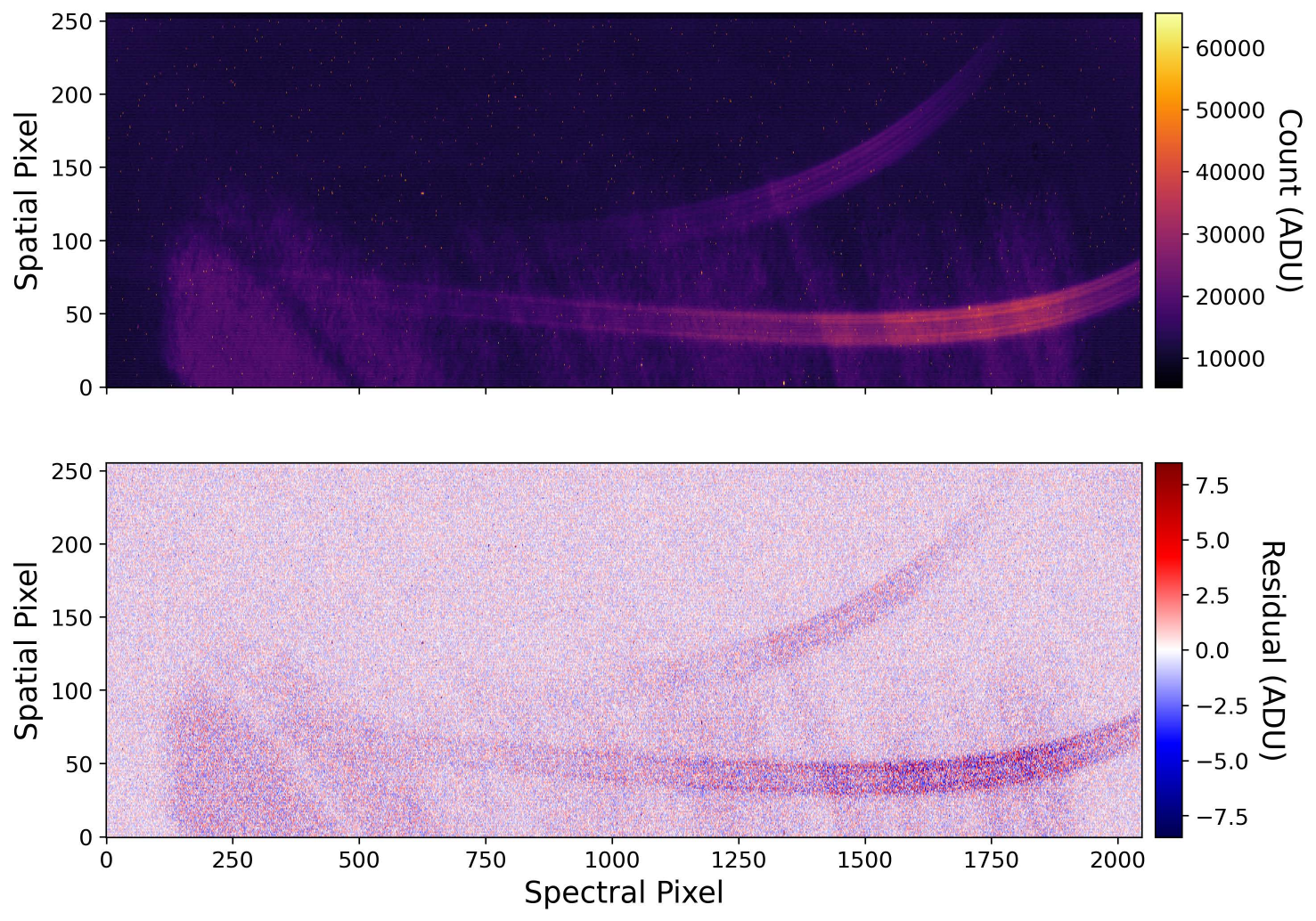} \\
    \vspace{0.0cm} 
    \includegraphics[width=0.9\linewidth]{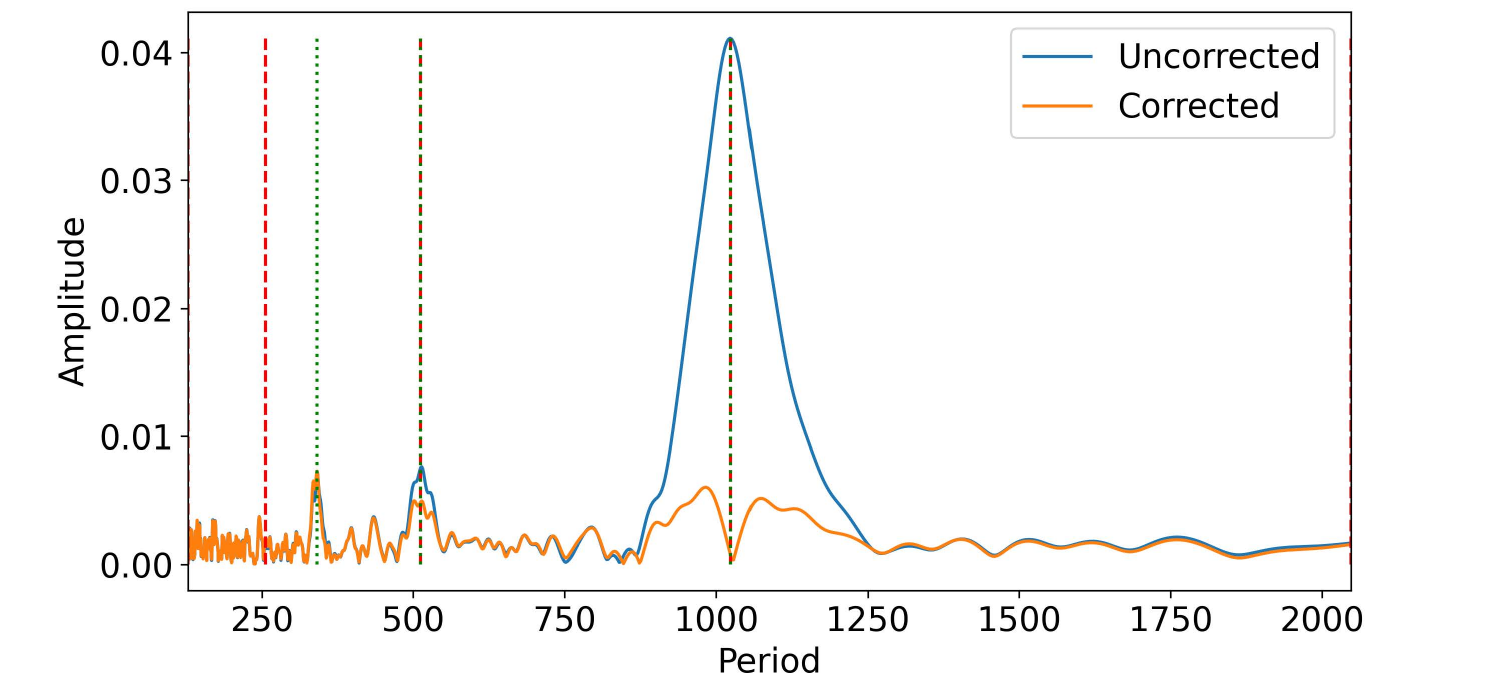}
    \caption{Raw NIRISS/SOSS spectrum (top) and difference between the corrected and uncorrected raw spectra (middle) from the first integration for WASP-39\,b. Due to our model of the periodic \ac{inl} as a sinusoid with an amplitude proportional to the signal level, the effects of the correction are stronger in high signal regions of the image, including the spectral trace and structure in the pedestal and bias level. A \ac{ls} of each pixel's ramp fit residuals (bottom), similar to bottom right panel in Figure~\ref{fig:adcinl} but for the WASP-39\,b ERS data, shows much of the power in the dominant 1024 period has been successfully calibrated out.}
    \label{fig:wasp39nirisstrace}
\end{figure*}

The result of this correction, as well as the before-and-after \ac{ls} is shown in Figure~\ref{fig:wasp39nirisstrace}. The correction removes nearly all of the 1024-period sinusoid from the \ac{ls} as before, validating that this is a consistent effect across datasets in \ac{niriss}. Because the amplitude of our sinusoidal model is taken to be proportional to the raw pixel count (see Section~\ref{sec:inl}, where we show that the amplitude of the sinusoid evolves with the signal level), higher signal pixels undergo a stronger correction (see middle panel of Figure~\ref{fig:wasp39nirisstrace}. 

\subsection{Data Reduction}
\label{sec:reduction}

\begin{figure*}
    \centering
    \includegraphics[width=\linewidth]{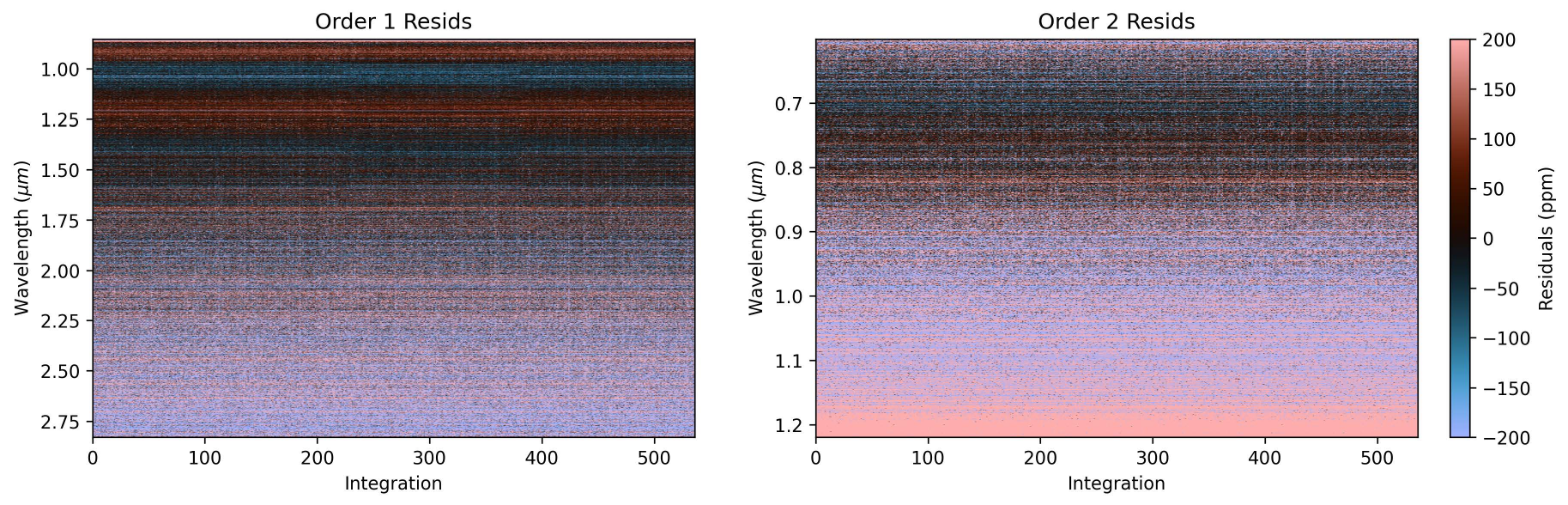}
     \caption{Residuals obtained from subtracting the corrected and uncorrected reduced spectra from the NIRISS SOSS data of WASP-39 (corrected-uncorrected). Individual pixels routinely exhibit more than 200\,ppm of difference between the corrected and uncorrected data. Different wavelengths have different average pixel counts; the periodic INL thus applies a correction to each wavelength at a different level. Because the SOSS PSF is relatively stable over time, the horizontal banding is most visible in the residual correction in both Order 1 and 2. The imprint of the transit is visible as more faint vertical striations between integrations 200-400. Some of the vertical banding is due to the flux-dependence of the periodic INL changing the transit depth, as the temporal reduction in flux puts the pixels at a different phase in the sinusoid.}
    \label{fig:1dorder1resids}
\end{figure*}

We reduce the \ac{adc} \ac{inl} corrected and the uncorrected data identically using the ExoTEDRF (formerly supreme-SPOON) package from the Stage~0 data \citep{feinstein2023, radica2023, radica2024}. We apply the Stage 1-3 workflow exactly as described in \citet{radica2023}, which we briefly cover here. We first apply detector-level corrections: data-quality initialization, saturation flagging, superbias subtraction, reference-pixel correction, dark-current subtraction, linearity correction, cosmic-ray/jump detection, ramp fitting, and gain scaling. We then correct the structured SOSS background and group-level column-correlated $1/f$ noise, using the first 150 and final 100 integrations as the out-of-transit baseline. We correct remaining bad pixels and temporal outliers with spatial and temporal filtering. Detector-correlated systematics are corrected by reconstructing the time series after removing PCA components 2--4, which are caused by detector noise and not astrophysical variability. The spectral trace was measured directly from the data, zeroth order contaminants were masked using the accompanying F277W exposure, and the final spectra were extracted using a 30-pixel-wide box aperture centered on the measured trace positions. The extracted spectra were normalized by the median out-of-transit baseline to produce the 1D spectra that will be used for transit fitting. The difference between the identically reduced corrected and uncorrected spectra are shown in Figure~\ref{fig:1dorder1resids} for Order 1 and Order 2.

\subsection{Transit Fitting} \label{sec:transit}

To fit the transit depths at each wavelength bin in NIRISS/SOSS we use a custom Python script based on the \texttt{jaxoplanet} code \citep{jaxoplanet}. We define the priors and likelihood using the probabilistic programming language (PPL) \texttt{numpyro} \citep{phan2019}. In our model the transit parameters specific to the orbit, such as the mid-transit time, transit duration and impact parameter are fit but held constant across wavelengths. The orbital period is held fixed at the value reported in \citet{faedi2011} and the orbit is assumed to be circular. We define the transit depth, quadratic limb darkening coefficients, an offset term, and a jitter parameter to be fit at each wavelength independently. We bin the spectrum into 128 wavelength bins in Order 1 and 61 bins in Order 2. We use \texttt{vmap} in \textsc{Jax} to vectorize the transit fit across wavelength for the parameters specified above for computational efficiency.

We applied tight normal priors to the mid-transit time, $\mathcal{N}(59787.056726, 10^{-3})$ MJD and a lognormal prior on the transit duration in hours, $\mathcal{N}(\ln(2.8032), 10^{-3})$. The impact parameter was drawn from a truncated normal prior, $\mathcal{N}(0.4530, 0.1)$, bounded between 0.3 and 0.5. For each independent wavelength bin, the transit depths were fitted using a truncated normal prior bounded between 0 and 1, $\mathcal{N}(0.14602, 10^{-2})$, from which the planet-to-star radius ratios were deterministically derived. Quadratic limb-darkening coefficients were parameterized as offsets from theoretical ExoCTK predictions \citep{stevenson2018}, with a zero-mean normal prior $\mathcal{N}(0.0, 0.1)$ placed on these deviations. Baseline flux offsets for each bin were constrained using a tight normal prior, $\mathcal{N}(0.0, 10^{-4})$. The total variance for each wavelength bin was modeled as the quadrature sum of the reported observational uncertainties and an independent white noise jitter term, which was drawn from a half-normal prior, $\mathcal{HN}(10^{-4})$. We use the \ac{nuts} sampler \citep{hoffman2011} with Hamiltonian Monte Carlo to obtain posterior distributions for the transit parameters and transmission spectrum. We start by optimizing the model to find the maximum likelihood solution using the \texttt{numpyro-ext} package, a wrapper of \textsc{Jax}\texttt{opt} \citep{jaxopt_implicit_diff}. We then initialize 4 chains at the maximum likelihood solution with 3000 steps for warm-up before sampling 7000 steps in each chain. This entire process, including the priors, likelihood, optimization and posterior inference are applied identically on the corrected and uncorrected reduced spectra.  

\begin{figure*}
    \centering
    \includegraphics[width=\linewidth]{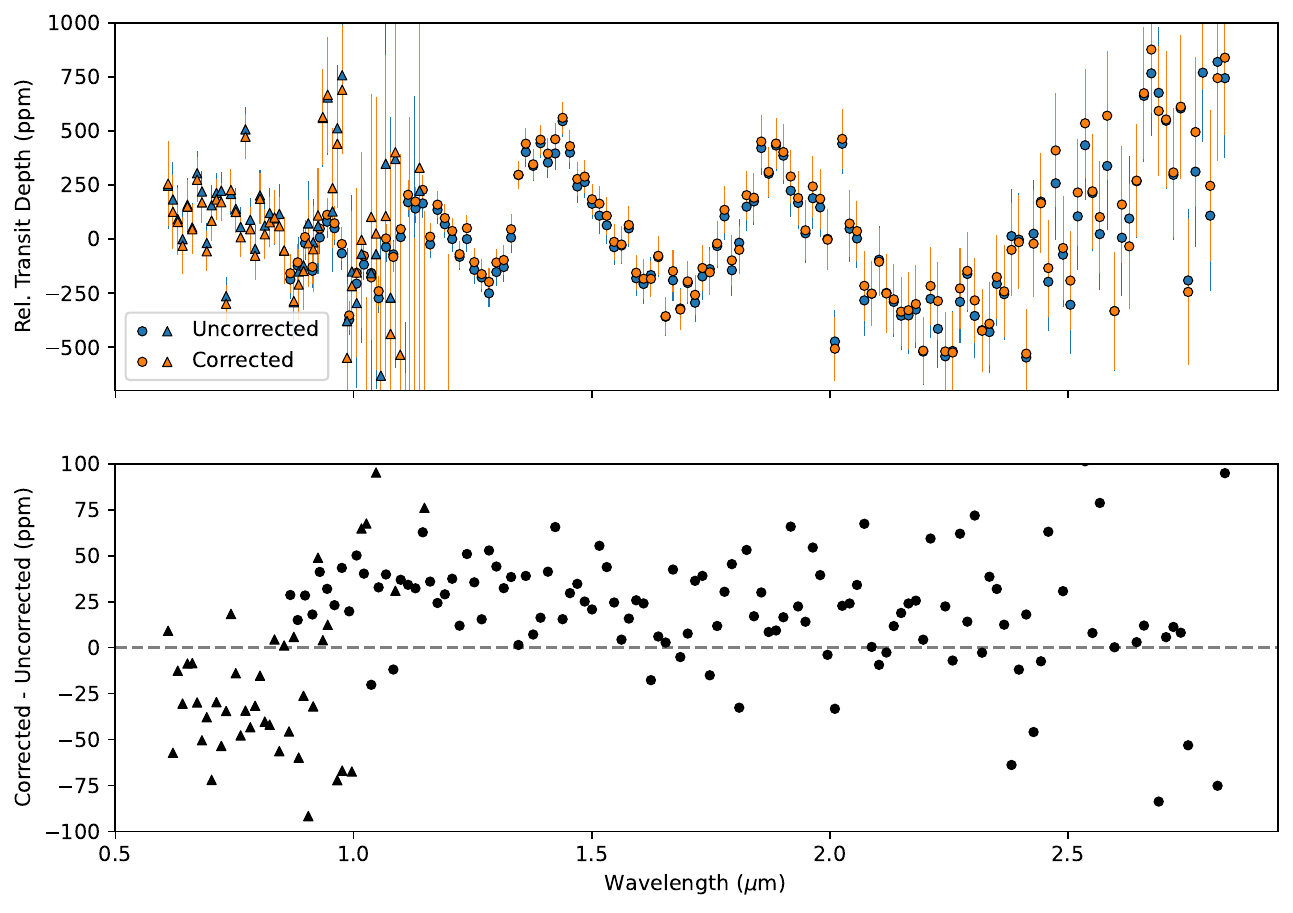}
    \caption{Transmission spectrum of WASP-39\,b with (orange) and without (blue) correction of the inferred periodic integral nonlinearity. With the new correction, the median transit depth is deeper by 25\,ppm across Order 1 (circles) and shallower by 30\,ppm across Order 2 (triangles). The transit depths exhibit dispersions of 25\,ppm and 59\,ppm in Orders 1 and 2 respectively, indicating that the periodic \ac{inl} imparts a deterministic and systematic noise contribution to transmission spectra.}
    \label{fig:transmissionspectrum}
\end{figure*}

The results of the transit fits are displayed in Figure~\ref{fig:transmissionspectrum}. Across both spectral orders, we observe significant differences between the corrected and uncorrected transmission spectra. The median absolute deviations between the corrected and uncorrected transmission spectra are 17\,ppm and 40\,ppm in Orders 1 and 2 respectively. In addition, we find that the correction imparts an additive offset of 25\,ppm in Order 1 and -30\,ppm in Order 2. No statistically significant slopes are found in the residuals for either Order 1 or Order 2. Taking the \ac{ls} of the transmission spectrum residuals, we also do not see any statistically significant evidence of oscillations in the transmission spectrum induced by the periodic \ac{inl}.

We also can demonstrate the significance or bias that the correction imparts on the spectrum using the z-score, or more generally the Mahalanobis distance \citep{mahalanobis1936} as follows: we take the mean and standard deviation of the corrected transmission spectrum and measure the statistical distance from the mean uncorrected spectrum. In effect, we are aggregating the systematic offset of each data point in the transmission spectrum relative to its uncertainty. We find that the correction shifts the transmission spectrum by $4.25\,\sigma$ in Order~1 and $2.61\,\sigma$ in Order~2, and taken as a whole, represents a $5.0\,\sigma$ distance from the uncorrected spectrum. We discuss the origin of these offsets and their implications further in Section~\ref{sec:discussion}. 

\begin{figure*}
    \centering
    \includegraphics[width=0.83\linewidth]{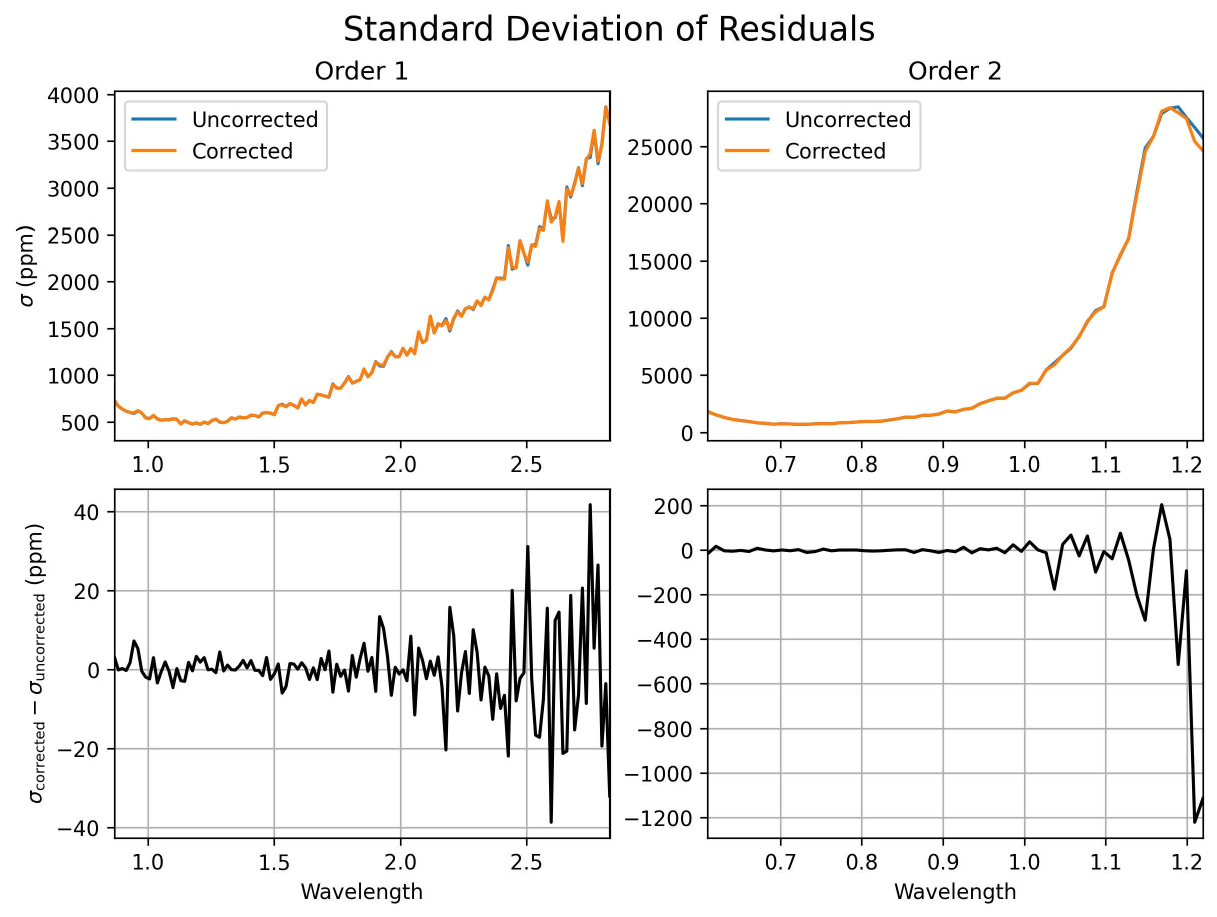}
    \caption{Standard deviation of transit model residuals as a function of wavelength for the periodic \ac{inl}-corrected and uncorrected datasets. Bottom panel shows the difference in standard deviations; positive and negative values indicate the correction worsens or improves the scatter respectively.}
    \label{fig:residualsigma}
\end{figure*}

We also analyzed the residuals and noise statistics to assess whether the periodic \ac{inl} also results in some stochastic noise source. We find no significant difference in the standard deviation of the transit model residuals as a function of wavelength for either Order 1 or Order 2, demonstrated in Figure~\ref{fig:residualsigma}. We also analyzed the posterior  jitter term, added in quadrature to the error bars for each wavelength bin in our transit fit, to evaluate whether any sources of error that were unaccounted for in the exoTEDRF pipeline were different between the corrected and uncorrected data. We found no difference, with both corrected and uncorrected data yielding a posterior jitter of $75\pm54$ ppm. 

\section{Discussion}
\label{sec:discussion}


We find a significant overall effect of correcting the periodic integral nonlinearity effect in NIRISS data. This correction manifests in two primary ways: changing the depth of the transit across orders and dispersing the transit depths from their uncorrected values. 

WASP-39\,b's ERS dataset is an example of temporal trends in flux being systematically offset by the periodic \ac{inl}. Similarly, spatial variations of close to 512 counts or odd integer multiples thereof can cause equally large deviations from ideal linear response, as originally noticed in \ac{niriss} \ac{ami}  \citep{amigo2025}. These effects are all systematic in nature, not stochastic, unlike $1/f$ noise or other sources of random error. We were not able to find obvious signs of a stochastic noise floor added by the periodic \ac{inl}, with the standard deviations of residuals, flux errors estimated by the exoTEDRF pipeline, and an additive jitter term showing no significant differences between the corrected and uncorrected data. 

Many of the results we present in Section~\ref{sec:wasp39} relating to the correction and its effect on transmission spectroscopy in the \ac{soss} mode of \ac{niriss} are based on our model assumptions for the  periodic \ac{inl}. While we show that the correction does remove most of the power in the dominant 1024-DN mode and the harmonics (see Figure~\ref{fig:wasp39nirisstrace}), the residuals show some periodic signals do remain, especially at the lower and higher signal ends. Our model for the periodic \ac{inl} as a multiplicative effect with an amplitude proportional to the raw pixel count may be the limiting factor in the accuracy of the correction. If the true \ac{inl} effect has a changing amplitude; for instance, an amplitude that is a more complex function of the signal level, our correction would not fully remove the effect. We briefly propose a method to achieve a more confident correction in Section~\ref{sec:conclusion}, but defer this to future works. 


\subsection{Transmission spectrum offsets}

We find that the correction of the periodic nonlinearity can either deepen or shallow the transit. For WASP-39\,b this effect is particularly instructive; the exoplanet transit reduces the data values in parts of the order 1 NIRISS trace by approximately 700 counts, relatively near half the dominant 1024-count period of the periodic \ac{inl}. Because of the flux-dependent nature of the periodic \ac{inl}, temporal changes in flux (such as due to a transit) can be either under or overestimated. Depending on the average ADU counts in the out-of-transit spectral trace, the periodic INL can modify the transit to be either deeper or shallower at levels up to the peak-to-peak amplitude of the effect, roughly 249\,ppm. This should be an upper limit; the finely-structured \ac{soss} PSF with a diversity of raw pixel counts within a wavelength bin should result in some averaging of this effect.

Furthermore, the average ADU count varies significantly between and within each spectral order, which can lead to spurious offsets between orders. In NIRISS data for WASP-39\,b, the region in Order 1 between $1.0-1.5\,\mu$m has an average raw pixel value of about 37k, whereas the region in Order 2 leading up to $1\,\mu$m has an average pixel count of about 20k. This leads to drastically different \ac{inl} corrections for these two orders. Tracing the average correction for $2.1\%$ transit dip within each of these segments, we find that the periodic \ac{inl} should reduce and increase the transit depth in these regions in Order 1 and 2 respectively. We show in Section~\ref{sec:transit} that we see exactly this direction of offsets in the transmission spectrum induced by the periodic \ac{inl} resulting in a significant total offset between the two orders of 55\,ppm. 

Multiple recent works have explored such offsets between instruments and orders, including between NRS1 and NRS2 in NIRSpec \citep{moran2023, wallack2026} as well as between NIRSpec and NIRISS \citep{madhusudhan2023, carter2024, schmidt2025}. \citet{carter2024} found a 138\,ppm shift in an overlapping region between NIRSpec G395H and NIRISS/SOSS data on WASP-39~b, which could partially be explained by the up to 249\,ppm peak-to-peak shift induced by the periodic \ac{inl} on the transit depths in our analysis. Significant attention has also been devoted to these offsets in the interpretation of potential biosignatures in transmission spectra as these terms could cause degeneracies in retrievals \citep{madhusudhan2023, schmidt2025}. In many such analyses, including packages to perform transmission spectroscopy \citep{parvianen2025}, an offset parameter is assigned to each order and instrument to account for these shifts. 

The average raw flux within each order also varies, with the trace smoothly varying in intensity over wavelength. This could cause spurious slopes or oscillations in the transmission spectrum depending on the flux change across each trace. While we were not able to find statistically significant evidence of these effects in the WASP-39\,b ERS data, we nevertheless caution that they are possible and may be present in other datasets. 

\newpage
\section{Conclusions}
\label{sec:conclusion}

In this work, we use archival, raw data from JWST \ac{niriss} to establish the existence of a periodic \acl{inl} imparting a flux-dependent sinusoidal systematic in data taken with the \ac{niriss} instrument. We measure the periodic \ac{inl}'s properties and derive a correction, finding a peak-to-peak amplitude of 249\,ppm.

This is not likely to be the dominant limiting factor in the performance of \ac{niriss}, but as shown in Section~\ref{sec:wasp39} it presents a $5-\sigma$ shift in the recovered transmission spectrum and in its interpretation. Specifically, we note that the correction of the periodic \ac{inl} removes a $\sim$ 55\,ppm offset between \ac{niriss} Orders 1 and 2, and potentially also may explain shifts of the entire \ac{niriss} transmission spectrum relative to other instruments. 

While our analysis confirms one particular cause for such an offset in \ac{niriss}, such offsets have also been observed in other instruments such as NIRSpec where we see no evidence for a periodic \ac{inl}. Effects such as starspots could be responsible for some offsets between instruments for data taken at different times. Other detector nonlinearities or effects could also be responsible for these offsets and more work is needed to identify their causes especially given their importance in the joint interpretation of spectra across orders and instruments in JWST. Until the causes for these shifts can be identified across JWST instruments and corrected, our work underscores the importance of adding an offset parameter for each order and instrument in analyses of exoplanet transmission spectra. 

While we have offered a Python implementation of our best calibration for this \ac{inl}, it is beyond the resources available to our research group to optimize this further due to the memory and compute difficulties of fitting large numbers of pixels simultaneously. The analysis presented here is limited by the small numbers of pixels in our training data recording very high or low fluxes. The user community may benefit from loading potentially orders of magnitude more data into the bag-of-pixels approach we have taken, and deriving still more accurate \ac{inl} including better performance at the high and low count end. 
Rather than an iterative fit, with sufficient access to computing resources it may be better to take a hierarchical Bayesian approach to forward-model these effects over multiple datasets and quantify the uncertainty in the parameters. The assumption of an amplitude proportional to the count level can also be relaxed, for instance by using a Gaussian process (GP) model for the \ac{inl}, allowing for a more precise calibration. This might be efficiently implemented in a probabilistic programming language such as \textsc{NumPyro} \citep{phan2019}. 

As the upcoming \textit{Nancy Grace Roman Space Telescope} \citep{roman} and \textit{Lazuli Space Observatory} \citep{lazuli} both use H4RG detectors with similar readout electronics in some or all of their instruments, it will be important to do as much testing and calibration of these \acp{adc} on the ground as is possible before launch; and it may be necessary to conduct post-launch calibration of the \ac{inl} in the manner demonstrated in this paper to achieve final accurate corrections.

\section{Code and Data Availability}
\label{sec:data}

All data used to produce this work are publicly available. An \ac{inl} correction and pipeline is provided open source on GitHub at  \href{https://github.com/shashankdholakia/niriss-cal-inl}{https://github.com/shashankdholakia/niriss-cal-inl}. All data in the paper are available from the \ac{mast} under the appropriate proposal numbers.

\begin{acknowledgments}

We are grateful to Markus Loose, Jim Beletic, Eddie Bergeron, and Chelsea Huang for conversations about this paper. We are grateful in particular for the support of the Space Telescope Science Institute for their extensive work in understanding, commissioning, and supporting JWST and its instruments, including significant correspondence in this paper and our related works.

Shashank Dholakia and Benjamin Pope were funded by the Australian Government through the Australian Research Council DECRA fellowship DE210101639, and Shrish Ray and Pope by DP230101439. We are grateful to the Australian public for enabling this science. BP and SR would like to thank the Big~Questions~Institute for their philanthropic support.

This work was supported by resources provided by the University of Queensland Research Computing Centre’s \href{https://dx.doi.org/10.48610/wf6c-qy55}{Bunya supercomputer}, with funding from the University of Queensland, Brisbane, Australia.

We acknowledge and pay respect to the traditional owners of the land on which the University of Queensland, University of Southern Queensland, University of Sydney, and Macquarie University are situated, upon whose unceded, sovereign, ancestral lands we work. We pay respects to their Ancestors and descendants, who continue cultural and spiritual connections to Country. 

\end{acknowledgments}

\begin{contribution}

The ADC integral nonlinearity was first identified by LD and explained by BJSP and PT in the course of AMI calibration. AS identified further datasets to examine and advised on the instrument architecture and pipelines. SD1 and SR then identified the effect in SOSS and SD1 derived calibrations. SD2 reduced the corrected and uncorrected data and SD1 and SD2 analyzed effects on transmission spectroscopy. BJSP supervised the project. All authors contributed to the text.

\end{contribution}


\facilities{JWST (NIRISS)}

\software{
Astropy \citep{astropy:2022}, \textsc{NumPy} \citep{numpy}; Matplotlib \citep{matplotlib}; \textsc{Jax} \cite{jax}; \texttt{scipy} \citep{scipy}; \texttt{jaxoplanet} \citep{jaxoplanet}; \textsc{NumPyro} \citep{phan2019};
}




\newpage
\bibliography{ms}{}
\bibliographystyle{aasjournalv7}



\end{document}